\documentclass[aps,prl,twocolumn,superscriptaddress,showpacs,floatfix]{revtex4-1}
\usepackage[english]{babel}
\usepackage{graphicx}
\usepackage{bm}
\usepackage{stmaryrd}
\usepackage{amssymb}
\usepackage{amsfonts}
\usepackage{amsmath}
\usepackage{color}
\usepackage{xspace}

\DeclareGraphicsExtensions{.pdf,.ps,.eps}

\newcommand{\ket}[1]{\ensuremath{|#1\rangle}\xspace}

\begin{document}

\title{Electron spin resonance shift in spin ladder compounds}
\author{Shunsuke C. Furuya}
\affiliation{Institute for Solid State Physics, University of Tokyo, Kashiwa 277-8581, Japan}
\author{Pierre Bouillot}
\affiliation{DPMC-MaNEP, University of Geneva, CH-1211 Geneva, Switzerland}
\author{Corinna Kollath}
\affiliation{D\'epartement de Physique
Th\'eorique, University of Geneva, CH-1211 Geneva, Switzerland}
\affiliation{Centre de Physique Th\'eorique, Ecole Polytechnique, CNRS, 91128
 Palaiseau Cedex, France}
\author{Masaki Oshikawa}
\affiliation{Institute for Solid State Physics, University of Tokyo, Kashiwa 277-8581, Japan}
\author{Thierry Giamarchi}
\affiliation{DPMC-MaNEP, University of Geneva, CH-1211 Geneva, Switzerland}

\date{\today}

\begin{abstract}
We analyze the effects of different coupling anisotropies in spin-$1/2$ ladder on the Electron Spin Resonance (ESR) shift. Combining a perturbative expression in the anisotropies with temperature dependent Density Matrix Renormalization Group (T-DMRG) computation of the short range correlations, we provide the full temperature and magnetic field evolution of the ESR paramagnetic shift. We show that for well chosen parameters the ESR shift can be in principle used to extract quantitatively the anisotropies and, as an example, discuss the material BPCB.
\end{abstract}

\maketitle
Quantum magnetism is a fascinating branch of solid state physics \cite{Auerbach_book_magnetism}, and more recently of quantum gases in optical lattices \cite{bloch_cold_atoms_optical_lattices_review}. Due to the competition between the
exchange interactions between spins localized on different lattice sites, many fascinating quantum phases can occur ranging from ferromagnetically or antiferromagnetically
ordered states to spin liquids. Recently spin systems have also allowed one to address questions usually related to itinerant quantum 
systems such as Bose-Einstein condensation (BEC)~\cite{giamarchi_BEC} and Tomonaga-Luttinger liquid physics~\cite{Klanjsek_NMR_3Dladder}. 

For potential application it is particularly important to have efficient probes to characterize the properties of magnetic systems. Fortunately powerful probes, such as neutron scattering~\cite{lovesey_neutron_scattering} or nuclear magnetic resonance~\cite{slichter_NMR_book} exist and give important physical information. However in systems which are particularly sensitive to the symmetry of exchange interactions (such for example 
the ones where BEC can be realized), a direct probe of small asymmetries of the coupling constants is desirable. An ideal probe for this purpose is the Electronic Spin Resonance (ESR) which allows one
to investigate the frequency shift of the paramagnetic
resonance induced by small anisotropies~\cite{KanamoriTachiki,NagataTazuke,Maeda_perturbation}. In order to interpret the experimental observations 
an efficient theoretical analysis of the corresponding spectra is mandatory. Although theoretical interpretations have been developed for a long time~\cite{KuboTomita,mori_esr,KanamoriTachiki,NagataTazuke}, their range of applicability is generally limited to several models in a restricted domain of parameters: low temperature region of spin-$1/2$ antiferromagnetic
chains~\cite{oshikawa_esr_spinchains}, high temperature limit of spin-$1/2$ antiferromagnetic chains~\cite{ElShawish},
low temperature region of spin-1 Haldane chain compound~\cite{affleck_esr_NENP,sakai_NENP}, spin-1/2 chains with nearest
and next nearest neighbor interactions~\cite{zvyagin_esr}, strongly dimerized spin-$1/2$ ladders analysed in a single rung picture~\cite{cizmar_esr_bpcb}.

In this letter we present a flexible theory of the ESR spectra.  Combining an analytical expression of the ESR shift with a numerical evaluation of short range correlations we give the full dependence of the ESR shift with the magnetic field and the temperature. We apply this approach to the case of spin-$1/2$ ladders using
 temperature dependent density matrix renormalization group (T-DMRG)~\cite{White_finT,Bouillot_ladder_statics_dynamics}. We show in particular that the ESR technique can be used to distinguish the nature of different anisotropies quantitatively. We illustrate our theory by applying it to ESR measurements on the compound $\mathrm{(C_5H_{12}N)_2CuBr_4}$ (BPCB) that have been performed~\cite{cizmar_esr_bpcb} recently.  Our study extends the analysis of the ESR measurements from Ref.~\cite{cizmar_esr_bpcb} focussing on the temperature and magnetic field dependence of the shift of the paramagnetic peak. We discuss the nature of different anisotropies deduced for BPCB and make proposals for additional experiments.

A spin-$1/2$ ladder in a magnetic field with interaction exchange anisotropies can be described by the Hamiltonian $\mathcal H = \mathcal H^0 +\mathcal H'
  \label{eq:H}$ which consists of two parts.
The first part
\begin{equation}
 \mathcal H^0 = J_\parallel \sum_{i,l}  {\bm S}_{i,l}
 \cdot {\bm S}_{i+1, l} + J_\perp \sum_i {\bm S}_{i,1} \cdot {\bm S}_{i,2}- g \mu_Bh S^z
 \label{eq:H0}
\end{equation}
includes the isotropic exchange interaction and the Zeeman term. The spin operator is defined by ${\bm S}_{i,l}=(S^x_{i,l},S^y_{i,l},S^z_{i,l})={\bm \sigma}$, where $\bm \sigma$ is the Pauli matrix acting on the spin located on leg $l = 1,2$ and rung $i$, and ${\bm S}= \sum_{i,l} {\bm S}_{i,l}$ is the total spin. Here $J_\parallel,J_\perp>0$ are the antiferromagnetic couplings along the legs and the rungs of the ladder, $g$ is the Land\'e factor, $\mu_B$ is the Bohr magneton and $h$ is the strength of the static magnetic field. Note that by this $g\mu_Bh$ has the unit of energy.

The anisotropic exchange interaction
\begin{equation}
 \mathcal H' =  \sum_{p,i,l}
 J'_{\parallel,p} S^p_{i,l} S^p_{i+1, l} + \sum_{p, i} J'_{\perp, p}
 S^p_{i,1} S^p_{i,2}
 \label{eq:H'}
\end{equation}
contains different spatial anisotropies $J'_{\parallel, p}$ and $J'_{\perp, p}$ in the exchange interactions along the legs and the rungs, respectively. $p = a,b,c$ represents the principal axis coordinates ${\bm a}_\alpha,{\bm b}_\alpha,{\bm c}_\alpha$ of more general interactions $A_\alpha^{qq'}S^q_{i,l}S^{q'}_{j,k}$ instead of $J'_{\alpha,p}S^p_{i,l}S^p_{j,k}$ in~\eqref{eq:H'} (for dipolar interaction typically) with $A_\alpha^{qq'}$, $\alpha=\parallel,\perp$, are symmetrical tensors in every coordinate basis $\bm x,\bm y,\bm z$ (denoted $q,q'=x,y,z$) aligned with the magnetic field ${\bm h }= h {\bm z}$. In the ${\bm a}_\alpha,{\bm b}_\alpha,{\bm c}_\alpha$ coordinates, one denotes ${\bm z}=(z_{\alpha, a}, z_{\alpha, b},z_{\alpha, c})$.

The ESR frequency shift $\delta \omega$ is defined as the difference between the ESR paramagnetic frequency resonance, and the 
Zeeman frequency:
\begin{equation}
 \hbar\delta \omega = \hbar\omega_{\mathrm r} - H.
  \label{eq:shift}
\end{equation}
Here $\omega_{\mathrm r}$ is the resonance frequency and $H=g\mu_B h$. Assuming that the anisotropies are small, i.e. $|J'_{\parallel, p}|, |J'_{\perp, p}| \ll J_\parallel,J_\perp$, we determine~\eqref{eq:shift} based on a first order perturbation theory in the anisotropy. At this order, the ESR shift is represented by a static correlation function~\cite{KanamoriTachiki,NagataTazuke,Maeda_perturbation},
  \begin{equation}
  \hbar\delta \omega = - \frac{\langle [[\mathcal H', S^+],
    S^-] \rangle_0}{2\langle S^z \rangle_0},
    \label{eq:dw}
  \end{equation}
with $S^\pm = S^x\pm i S^y$. The average $\langle \cdot \rangle_0$ is taken with respect to
the unperturbed Hamiltonian $\mathcal{H}^0$~\eqref{eq:H0}.
The formula \eqref{eq:dw} holds for arbitrary temperatures $T$
and magnetic fields. For the spin ladder, the shift \eqref{eq:dw} can be decomposed in two components
\begin{equation}
  \hbar\delta \omega =  \sum_{\alpha = \parallel, \perp}
 f_\alpha(\bm z)\ Y_\alpha(T,H).
 \label{eq:dw_fY}
\end{equation}
corresponding to the effect induced by the anisotropy in the parallel ($\alpha=\parallel$) and the perpendicular ($\alpha=\perp$) interaction exchange. The temperature and magnetic field dependence of the shift \eqref{eq:dw_fY} 
are determined by the short range correlations $Y_\alpha(T,H)$:
\begin{align}
 Y_\parallel (T,H) &= \frac{\langle S^z_{i,1} S^z_{i+1,1} -
 S^x_{i,1}S^x_{i+1,1}\rangle_0}
 {\langle S^z_{i,1} \rangle_0},
 \label{eq:Y_parallel} \\
 Y_\perp (T,H) &= \frac{\langle S^z_{i,1} S^z_{i,2} - S^x_{i,1}
 S^x_{i,2}\rangle_0}{\langle S^z_{i,1} \rangle_0}.
 \label{eq:Y_perp}
\end{align}
Computing these correlations is in general involved~\cite{azzouz_short_correlation_ladder}. Here we use the T-DMRG method which allows us to determine the ESR shift. Note that this approach 
is quite general and can be extended to other systems. 

For the considered spin ladder we fix the direction of the magnetic field $\bm z$ with respect to the system orientation,
the factors $f_\alpha(\bm z)$ depend only on
the anisotropic coupling constants $J'_{\alpha, p}$:
\begin{equation}
  f_\alpha(\bm z) = t_\alpha \sum_{p = a,b,c} J'_{\alpha, p}
 (1-3{z_{\alpha, p}}^2),
 \label{eq:f}
\end{equation}
with $t_\parallel=1$ and $t_\perp=1/2$. Assuming that all the isotropic coupling is included in $\mathcal H^0$, i.e. $\sum_p J'_{\alpha,p}=0$, we obtain
$f_\alpha(\bm z)=-3t_\alpha\ \bm zA_\alpha\bm z^T$. Hence $f_\alpha(\bm z)$ measures the anisotropy $A_\alpha^{qq'}$ along the magnetic field orientation $\bm z$ and has extrema when $\bm z\parallel{\bm a}_\alpha,{\bm b}_\alpha$ or ${\bm c}_\alpha$.

\begin{figure}[h!]
 \centering
 \includegraphics[width=\linewidth]{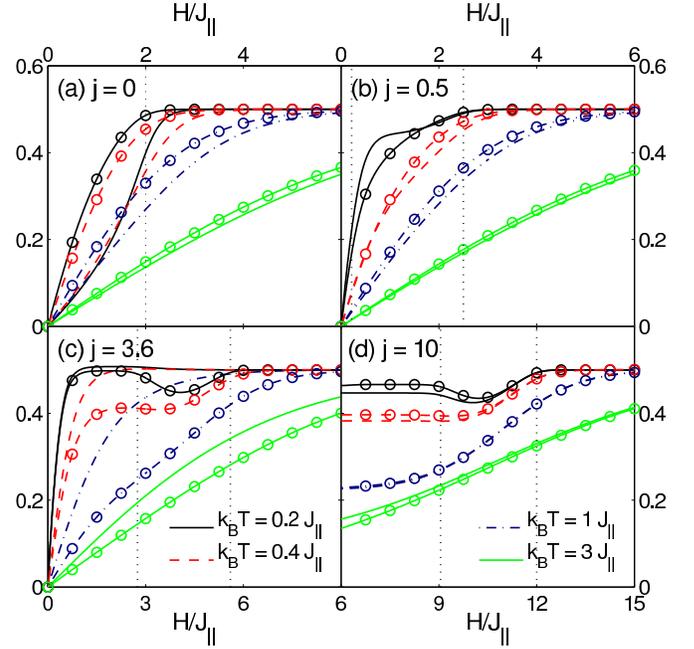}
 \caption{(color online): Magnetic field dependence of
$Y_\parallel$ (lines with circles) for several temperatures and coupling ratios $j$.  In (a-c), $Y_\perp$ is represented by lines without circles. In (d), the strong coupling $j\rightarrow\infty$ approximation~\eqref{eq:Y0_Y1_sc} of $Y_\parallel$ is shown with lines without circles.  For the BPCB coupling ratio $j=3.6$ in (c), the plotted temperatures	 are $T\approx0.72\ \text{K}$, $T\approx1.44\ \text{K}$, $T\approx3.6\ \text{K}$, $T\approx10.8\ \text{K}$ in experimental unit. The two vertical dotted lines locate the critical fields $H_{c1}$ and $H_{c2}$ respectively. \label{fig:Y0_Y1}}
\end{figure}

In the following we first discuss the properties of $Y_\alpha(T,H)$ for different coupling ratios $j=J_\parallel/J_\perp$. We then focus on the compound BPCB and analyze its ESR measurements.

At low magnetic field $H\ll H_{c1}$ both $Y_\parallel$ and $Y_\perp$ grow linearly
with $H$ (Fig.~\ref{fig:Y0_Y1}), i.e. $Y_\parallel,Y_\perp\propto H$ ($H_{c1}<H_{c2}$ are the two critical fields between which the system is gapless~\cite{Bouillot_ladder_statics_dynamics}). In a strong magnetic
field  $H\gg H_{c2}$ the system becomes totally polarized and 
$Y_\parallel,Y_\perp\approx 1/2$. Between these two
limits the behavior of $Y_\parallel$ and $Y_\perp$ is very subtle and clearly
depends on the coupling ratio $j$.
In the zero interladder coupling limit $j=0$, the two ladders decouple. $Y_\perp$ reduces to $Y_\perp=\langle S^z_{i,1}\rangle_0$, and $Y_\parallel$ can also be calculated analytically~\cite{Maeda_BetheAnsatz,oshikawa_esr_spinchains}. The system is gapless below $H_{c2}$ and
$Y_\parallel$, $Y_\perp$ are monotonic functions of $H$ as
shown in Fig.~\ref{fig:Y0_Y1}(a).

For finite interladder couplings $j>0$, the system is gapped for $H<H_{c1}$~\cite{chitra_spinchains_field} and has a ${\bm S}^2=S^z=0$ ground
state. Thus we expect that
$Y_\parallel$ and $Y_\perp$ show special signatures at $H_{c1}$ when the gap is closed and the excitations become
accessible. In this range of magnetic field $H\approx J_\perp$,
$Y_\parallel$ can exhibit a non-monotonical behavior at low temperatures and strong couplings ($j=3.6,10$ in Fig.~\ref{fig:Y0_Y1}(c-d)). In contrast $Y_\perp$ shows an intermediate plateau like feature at low coupling (cf.~$j=0.5$ in Fig.~\ref{fig:Y0_Y1}(b)) and a strong rise with a subsequent plateau at a value $0.5$ for stronger coupling.

The features at large $j$ are roughly understood using a strong coupling approach. In this limit for $H_{c1}<H<H_{c2}$ (gapless regime) and $k_BT\ll J_\perp$, the Hilbert space per rung can be approximated keeping only the two lowest states, i.e. the triplet $\ket{t^+}=\ket{\uparrow\uparrow}$ and singlet $\ket{s}=(\ket{\uparrow\downarrow}-\ket{\downarrow\uparrow})/\sqrt{2}$ state. Mapping these onto two effective spin-$1/2$ states $\ket{t^+}\rightarrow\ket{\tilde\uparrow}$ and $\ket{s}\rightarrow\ket{\tilde\downarrow}$, the system $\mathcal{H}^{0}$ becomes equivalent to an effective anisotropic spin chain~\cite{Tachiki_spin_ladder,chaboussant_ladder_strongcoupling,mila_ladder_strongcoupling,giamarchi_ladder_coupled}
$$
\mathcal{H}_{\text{XXZ}}=J_\parallel
\sum_{i}\left(\tilde S_i^x\tilde S_{i+1}^x+\tilde S_i^y\tilde S_{i+1}^y+\Delta\tilde S_i^z\tilde S_{i+1}^z\right)
-\tilde H\tilde S_T^z.
$$
with $\Delta=1/2$ and
$\tilde
H=H-J_\perp-J_\parallel/2$. Within this approximation
\begin{equation}\label{eq:Y0_Y1_sc}
Y_\perp=\frac{1}{2},\quad Y_\parallel=\frac{1/8+\langle \tilde S^z_i\rangle/2+\langle \tilde S^z_i\tilde S^z_{i+1}\rangle/2-\langle \tilde S^x_i\tilde S^x_{i+1}\rangle}{1/2+\langle \tilde S^z_i\rangle}.
\end{equation}
In particular, this approximation recovers the almost constant behavior of $Y_\perp$ in the gapless regime at low temperature. In contrast, although $Y_\parallel\approx1/2$ for $H\gg H_{c2}$, this quantity shows a dip in the gapless regime and a plateau when $0\ll H<H_{c1}$. These effects are present in Fig.~\ref{fig:Y0_Y1}(c,d) in both the strong coupling approximation and the full ladder at low temperature.
The use of Eq.~\eqref{eq:dw_fY} for the discrimination between the two kinds of anisotropies $J'_{\parallel, p}$ and $J'_{\perp, p}$ is very efficient in this range of parameters $H\approx J_\perp$ and $k_BT\approx J_\parallel$ for which very distinct behavior contributions from $Y_\parallel$ and $Y_\perp$ can be well separated. Note that when $j$ decreases these contributions become less distinguishable (see Fig.~\ref{fig:Y0_Y1}(b)).

\begin{figure}[h!]
 \centering
\includegraphics[width=\linewidth]{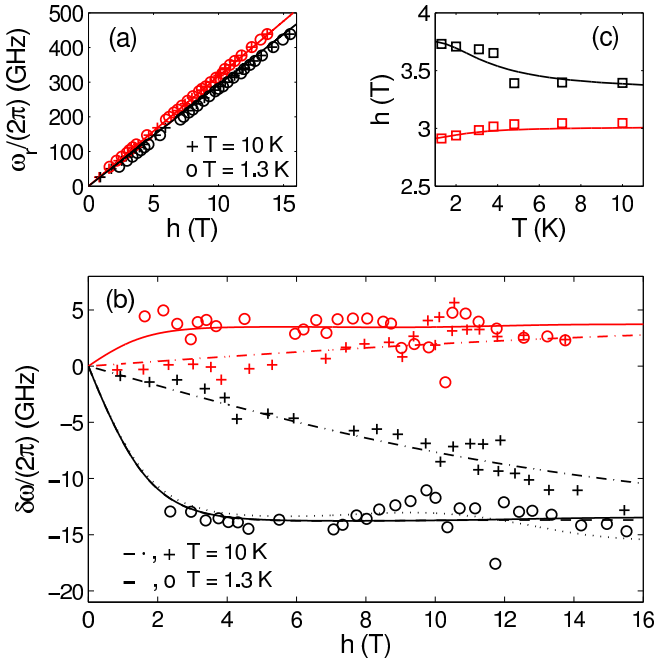}
\caption{(color online): Magnetic field dependence of (a) the ESR
 paramagnetic resonance frequency $\omega_r$, (b) the ESR resonance frequency shift $\delta\omega$. (c) Temperature dependence of the magnetic field at which the ESR paramagnetic resonance  at $\omega_r=2\pi\cdot96\
 \text{GHz}$ occurs. Symbols denote experimental data taken from~\cite{cizmar_esr_bpcb} and lines theoretical predictions of the best fitting
 parameters $f_{\parallel1}/k_B = 0.1 \ \text{K}$, $f_{\perp1}/k_B = -1.4\  \text{K}$ and $f_{\parallel2}/k_B = 0.2\  \text{K}$, $f_{\perp2}/k_B = 0.2\  \text{K}$ if not stated otherwise. The results corresponding
 to the two differently oriented ladders 1 and 2 are plotted in black and grey (online red), respectively. In (a) the linear Zeeman component of the
 ESR frequency resonance is represented by full lines (for the Land\'e
 factors $g_1$ and $g_2$).  We show in (b) the theoretical prediction of the ESR shift at $T=1.3\ \text{K}$ for the ladder 1 considering only the leg (black dotted lines) or the rung anisotropy (black dashed lines) with the fitting parameters~\eqref{eq:legparam} and~\eqref{eq:rungparam}, respectively. These fits show only small deviations with the best fit in this range of temperature and magnetic field.\label{fig:mpcorrelationmz}}
\end{figure}

Recently ESR measurements performed on BPCB~\cite{cizmar_esr_bpcb} have pointed out small anisotropies from the perfect Heisenberg Hamiltonian. BPCB is known to be a very good realization of weakly coupled spin-$1/2$ ladders with remarkably low couplings $J_\parallel/k_B = 3.6\ \text{K}$ and $J_\perp/k_B=12.9\ \text{K}$ ($j=3.6$) and has been
intensively investigated during the last decade with various experimental techniques~\cite{Patyal_BPCB,watson_bpcb,Klanjsek_NMR_3Dladder,Thielemann_ND_3Dladder,Thielemann_INS_ladder,Savici_BPCB_INS,Ruegg_thermo_ladder,Anfuso_BPCB_magnetostriction,lorenz_thermalexp_magnetostriction,cizmar_esr_bpcb,Bouillot_ladder_statics_dynamics}. This compound is composed of two differently oriented equivalent ladders that can be modelized with the Hamiltonian $\mathcal{H}$ with a very low interladder coupling of the order of 100 mK~\cite{Klanjsek_NMR_3Dladder,Thielemann_ND_3Dladder,Bouillot_ladder_statics_dynamics} which we neglect in the following.  Using the theoretical prediction of the ESR shift \eqref{eq:dw_fY}, we are able to extract an upper bound for the anisotropy along the legs and along the rungs assuming the limiting cases in which only $J'_{\parallel,p}$ and $J'_{\perp,p}$ are present, respectively. To do this we fit our results for the ESR shift~\eqref{eq:dw_fY} to the experimental data from~\cite{cizmar_esr_bpcb} (Fig.~\ref{fig:mpcorrelationmz}(a)) after substraction of the linear Zeeman component (Fig.~\ref{fig:mpcorrelationmz}(b)) with the Land\'e factors $g_1 = 2.09\pm0.01$ and $g_2 = 2.27\pm0.01$ (for the ladders 1 and 2 labeled as in Ref.~\cite{cizmar_esr_bpcb}). In these experiments the magnetic field is tilted by $45^\circ$ away from the $\bf b$ axis in the $\bf bc^\star$ plan ($\bf a$, $\bf b$, $\bf c$ are the unit cell vectors of BPCB).

Taking into account only the anisotropies along the legs $J'_{\parallel, p}$, the obtained fitting parameters are
\begin{equation}\label{eq:legparam}
f_{\parallel1}/k_B = -1.5\pm0.1\ \text{K}\quad,\quad
f_{\parallel2}/k_B = 0.4\pm0.1\ \text{K}
\end{equation}
for the corresponding orientation of each ladder respectively. Similarly, considering only anisotropies along the rungs $J'_{\perp, p}$ we get
\begin{equation}\label{eq:rungparam}
f_{\perp1}/k_B = -1.3 \pm 0.1\ \text{K}\quad,\quad
f_{\perp2}/k_B = 0.3 \pm 0.1\ \text{K}.
\end{equation}
Once extracted, these parameters determine the shift at all temperatures and values of the magnetic field. As shown in Fig.~\ref{fig:mpcorrelationmz}, both anisotropies are able to reproduce very well all the experimental measurements available~\cite{cizmar_esr_bpcb}. The discrepancies between the ESR measurements and the prediction~\eqref{eq:dw_fY} lie within the uncertainties due to extraction of the peak location from the measured ESR spectra. Indeed, the shift~\eqref{eq:dw_fY} is related to the first frequency momentum of the peak~\cite{Maeda_BetheAnsatz,oshikawa_esr_spinchains} instead of its maximum location which is not well defined when the peak is broadened. These discrepancies are clearly seen for temperatures $T\approx J_\parallel/k_B=3.6~\text{K}$  in Fig.~\ref{fig:mpcorrelationmz}(c) which shows the magnetic field location of the peak at $\omega_r=2\pi\cdot96\ \text{GHz}$ extracted from extrema of the measured ESR spectra (Fig.~4 of Ref.~\cite{cizmar_esr_bpcb}) versus the temperature. Let us note that the magnitude of the extracted parameters~\eqref{eq:rungparam} considering only rung anisotropy is in agreement with the anisotropy $f_{\perp1}/k_B=-1.1\ \text{K}$ and $f_{\perp2}/k_B=0.4\ \text{K}$ extracted in~\cite{cizmar_esr_bpcb} for a single rung model. However, taking the full ladder into account  enables in principle the determination of the repartition of the anisotropy along the rung and the leg using suited experimental measures. 

\begin{figure}[h!]
 \centering
\includegraphics[width=\linewidth]{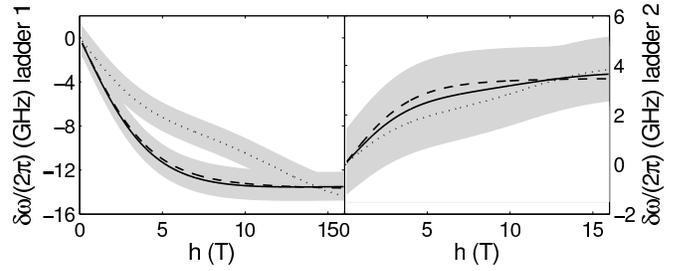}
\caption{Theoretical prediction of the ESR shift $\delta \omega$ versus the magnetic field at $T=3~\text{K}$ for the two ladders orientations. The  predictions are plotted in dotted, dashed and full lines considering only the rung, leg anisotropy with the fitting parameters~\eqref{eq:legparam},~\eqref{eq:rungparam} and with the best fitting parameters given in Fig.~\ref{fig:mpcorrelationmz}, respectively. The shaded area represents the standard deviation of the ESR experiments in Ref.~\cite{cizmar_esr_bpcb} from our theoretical prediction.  Note that for the experimental configuration of Ref.~\cite{cizmar_esr_bpcb}, the anisotropies along the magnetic field for ladder 1 could be well separated from the ESR measurements at $T=3~\text{K}$.\label{fig:ESR_prediction}}
\end{figure}

Since the measurements in Ref.~\cite{cizmar_esr_bpcb} are done in a region where $Y_\parallel$ and $Y_\perp$ are not very distinct, the fit cannot uniquely determine their contributions. The optimal fit is shown in Fig.~\ref{fig:mpcorrelationmz}(b,c). More precise measurements around $T\approx 3~\text{K}$, for which $Y_\parallel$ and $Y_\perp$ behave very distinctly (see Fig.~\ref{fig:Y0_Y1}(c)), could be used for a clear separation of the effects of both anisotropies. The differences expected at that temperature are shown in Fig.~\ref{fig:ESR_prediction}. Note that one should make a compromise between the separation of the two components $Y_\parallel$ and $Y_\perp$ and the occurrence of a well resolved ESR resonance peak. The extraction of the anisotropies $J_{\alpha,p}'$ and their corresponding orientation ${\bm a}_\alpha,{\bm b}_\alpha,{\bm c}_\alpha$ would require measurements for different magnetic field directions. 

In summary, we proposed a general theory to determine the frequency shift of ESR in spin-$1/2$ ladder compounds.
By taking advantage of the wide applicable scope of the perturbative formula
\eqref{eq:dw}, combined with numerical evaluation of the static correlation functions at arbitrary temperatures, 
we computed the shift in a wide range of temperature and magnetic field and showed how the ESR technique can distinguish the nature of different anisotropies. The predicted ESR shift is in very good agreement with experimental data for BPCB. Based on our results, we propose a new set of experiments to better resolve the different anisotropies.

We are grateful to the authors of Ref.~\onlinecite{cizmar_esr_bpcb} for providing us their experimental data. We would like to thank E. \v{C}i\v{z}m\'ar, M. Klanj\v{s}ek, Y. Maeda and C. Ruegg for fruitful discussions. This work is partly supported by Global COE Program ``the Physical
Sciences Frontier'' and KAKENHI No. 21540381 from MEXT, Japan, and the Swiss NSF, under MaNEP and Division II.


\begin{thebibliography}{34}%
\makeatletter
\providecommand \@ifxundefined [1]{%
 \@ifx{#1\undefined}
}%
\providecommand \@ifnum [1]{%
 \ifnum #1\expandafter \@firstoftwo
 \else \expandafter \@secondoftwo
 \fi
}%
\providecommand \@ifx [1]{%
 \ifx #1\expandafter \@firstoftwo
 \else \expandafter \@secondoftwo
 \fi
}%
\providecommand \natexlab [1]{#1}%
\providecommand \enquote  [1]{``#1''}%
\providecommand \bibnamefont  [1]{#1}%
\providecommand \bibfnamefont [1]{#1}%
\providecommand \citenamefont [1]{#1}%
\providecommand \href@noop [0]{\@secondoftwo}%
\providecommand \href [0]{\begingroup \@sanitize@url \@href}%
\providecommand \@href[1]{\@@startlink{#1}\@@href}%
\providecommand \@@href[1]{\endgroup#1\@@endlink}%
\providecommand \@sanitize@url [0]{\catcode `\\12\catcode `\$12\catcode
  `\&12\catcode `\#12\catcode `\^12\catcode `\_12\catcode `\%12\relax}%
\providecommand \@@startlink[1]{}%
\providecommand \@@endlink[0]{}%
\providecommand \url  [0]{\begingroup\@sanitize@url \@url }%
\providecommand \@url [1]{\endgroup\@href {#1}{\urlprefix }}%
\providecommand \urlprefix  [0]{URL }%
\providecommand \Eprint [0]{\href }%
\providecommand \doibase [0]{http://dx.doi.org/}%
\providecommand \selectlanguage [0]{\@gobble}%
\providecommand \bibinfo  [0]{\@secondoftwo}%
\providecommand \bibfield  [0]{\@secondoftwo}%
\providecommand \translation [1]{[#1]}%
\providecommand \BibitemOpen [0]{}%
\providecommand \bibitemStop [0]{}%
\providecommand \bibitemNoStop [0]{.\EOS\space}%
\providecommand \EOS [0]{\spacefactor3000\relax}%
\providecommand \BibitemShut  [1]{\csname bibitem#1\endcsname}%
\let\auto@bib@innerbib\@empty
%</preamble>
\bibitem [{\citenamefont {Auerbach}(1998)}]{Auerbach_book_magnetism}%
  \BibitemOpen
  \bibfield  {author} {\bibinfo {author} {\bibfnamefont {A.}~\bibnamefont
  {Auerbach}},\ }\href@noop {} {\emph {\bibinfo {title} {{Interacting Electrons
  and Quantum Magnetism}}}},\ \bibinfo {edition} {2nd}\ ed.\ (\bibinfo
  {publisher} {Springer},\ \bibinfo {address} {Berlin},\ \bibinfo {year}
  {1998})\BibitemShut {NoStop}%
\bibitem [{\citenamefont {Bloch}\ \emph {et~al.}(2008)\citenamefont {Bloch},
  \citenamefont {Dalibard},\ and\ \citenamefont
  {Zwerger}}]{bloch_cold_atoms_optical_lattices_review}%
  \BibitemOpen
  \bibfield  {author} {\bibinfo {author} {\bibfnamefont {I.}~\bibnamefont
  {Bloch}}, \bibinfo {author} {\bibfnamefont {J.}~\bibnamefont {Dalibard}}, \
  and\ \bibinfo {author} {\bibfnamefont {W.}~\bibnamefont {Zwerger}},\
  }\href@noop {} {\bibfield  {journal} {\bibinfo  {journal} {Rev. Mod. Phys.}\
  }\textbf {\bibinfo {volume} {80}},\ \bibinfo {pages} {885} (\bibinfo {year}
  {2008})}\BibitemShut {NoStop}%
\bibitem [{\citenamefont {Giamarchi}\ \emph {et~al.}(2008)\citenamefont
  {Giamarchi}, \citenamefont {Ruegg},\ and\ \citenamefont
  {Tchernyshyov}}]{giamarchi_BEC}%
  \BibitemOpen
  \bibfield  {author} {\bibinfo {author} {\bibfnamefont {T.}~\bibnamefont
  {Giamarchi}}, \bibinfo {author} {\bibfnamefont {C.}~\bibnamefont {Ruegg}}, \
  and\ \bibinfo {author} {\bibfnamefont {O.}~\bibnamefont {Tchernyshyov}},\
  }\href@noop {} {\bibfield  {journal} {\bibinfo  {journal} {Nature Physics}\
  }\textbf {\bibinfo {volume} {4}},\ \bibinfo {pages} {198} (\bibinfo {year}
  {2008})}\BibitemShut {NoStop}%
\bibitem [{\citenamefont {Klanj\v{s}ek}\ \emph {et~al.}(2008)\citenamefont
  {Klanj\v{s}ek}, \citenamefont {Mayaffre}, \citenamefont {Berthier},
  \citenamefont {Horvati\'{c}}, \citenamefont {Chiari}, \citenamefont
  {Piovesana}, \citenamefont {Bouillot}, \citenamefont {Kollath}, \citenamefont
  {Orignac}, \citenamefont {Citro},\ and\ \citenamefont
  {Giamarchi}}]{Klanjsek_NMR_3Dladder}%
  \BibitemOpen
  \bibfield  {author} {\bibinfo {author} {\bibfnamefont {M.}~\bibnamefont
  {Klanj\v{s}ek}}, \bibinfo {author} {\bibfnamefont {H.}~\bibnamefont
  {Mayaffre}}, \bibinfo {author} {\bibfnamefont {C.}~\bibnamefont {Berthier}},
  \bibinfo {author} {\bibfnamefont {M.}~\bibnamefont {Horvati\'{c}}}, \bibinfo
  {author} {\bibfnamefont {B.}~\bibnamefont {Chiari}}, \bibinfo {author}
  {\bibfnamefont {O.}~\bibnamefont {Piovesana}}, \bibinfo {author}
  {\bibfnamefont {P.}~\bibnamefont {Bouillot}}, \bibinfo {author}
  {\bibfnamefont {C.}~\bibnamefont {Kollath}}, \bibinfo {author} {\bibfnamefont
  {E.}~\bibnamefont {Orignac}}, \bibinfo {author} {\bibfnamefont
  {R.}~\bibnamefont {Citro}}, \ and\ \bibinfo {author} {\bibfnamefont
  {T.}~\bibnamefont {Giamarchi}},\ }\href@noop {} {\bibfield  {journal}
  {\bibinfo  {journal} {Phys. Rev. Lett.}\ }\textbf {\bibinfo {volume} {101}},\
  \bibinfo {pages} {137207} (\bibinfo {year} {2008})}\BibitemShut {NoStop}%
\bibitem [{\citenamefont {Lovesey}(1986)}]{lovesey_neutron_scattering}%
  \BibitemOpen
  \bibfield  {author} {\bibinfo {author} {\bibfnamefont {S.~W.}\ \bibnamefont
  {Lovesey}},\ }\href@noop {} {\emph {\bibinfo {title} {Theory of Neutron
  Scattering from Condensed Matter}}}\ (\bibinfo  {publisher} {Oxford Science
  Publications},\ \bibinfo {year} {1986})\BibitemShut {NoStop}%
\bibitem [{\citenamefont {Slichter}(1980)}]{slichter_NMR_book}%
  \BibitemOpen
  \bibfield  {author} {\bibinfo {author} {\bibfnamefont {C.~P.}\ \bibnamefont
  {Slichter}},\ }\href@noop {} {\emph {\bibinfo {title} {Principles of magnetic
  resonance}}}\ (\bibinfo  {publisher} {Springer-Verlag},\ \bibinfo {address}
  {Berlin Heidelberg New-York},\ \bibinfo {year} {1980})\BibitemShut {NoStop}%
\bibitem [{\citenamefont {Kanamori}\ and\ \citenamefont
  {Tachiki}(1962)}]{KanamoriTachiki}%
  \BibitemOpen
  \bibfield  {author} {\bibinfo {author} {\bibfnamefont {J.}~\bibnamefont
  {Kanamori}}\ and\ \bibinfo {author} {\bibfnamefont {M.}~\bibnamefont
  {Tachiki}},\ }\href {\doibase 10.1143/JPSJ.17.1384} {\bibfield  {journal}
  {\bibinfo  {journal} {J. Phys. Soc. Jpn.}\ }\textbf {\bibinfo {volume}
  {17}},\ \bibinfo {pages} {1384} (\bibinfo {year} {1962})}\BibitemShut
  {NoStop}%
\bibitem [{\citenamefont {Nagata}\ and\ \citenamefont
  {Tazuke}(1972)}]{NagataTazuke}%
  \BibitemOpen
  \bibfield  {author} {\bibinfo {author} {\bibfnamefont {K.}~\bibnamefont
  {Nagata}}\ and\ \bibinfo {author} {\bibfnamefont {Y.}~\bibnamefont
  {Tazuke}},\ }\href {\doibase 10.1143/JPSJ.32.337} {\bibfield  {journal}
  {\bibinfo  {journal} {J. Phys. Soc. Jpn.}\ }\textbf {\bibinfo {volume}
  {32}},\ \bibinfo {pages} {337} (\bibinfo {year} {1972})}\BibitemShut
  {NoStop}%
\bibitem [{\citenamefont {Maeda}\ and\ \citenamefont
  {Oshikawa}(2005)}]{Maeda_perturbation}%
  \BibitemOpen
  \bibfield  {author} {\bibinfo {author} {\bibfnamefont {Y.}~\bibnamefont
  {Maeda}}\ and\ \bibinfo {author} {\bibfnamefont {M.}~\bibnamefont
  {Oshikawa}},\ }\href {\doibase 10.1143/JPSJ.74.283} {\bibfield  {journal}
  {\bibinfo  {journal} {J. Phys. Soc. Jpn.}\ }\textbf {\bibinfo {volume}
  {74}},\ \bibinfo {pages} {283} (\bibinfo {year} {2005})}\BibitemShut
  {NoStop}%
\bibitem [{\citenamefont {Kubo}\ and\ \citenamefont
  {Tomita}(1954)}]{KuboTomita}%
  \BibitemOpen
  \bibfield  {author} {\bibinfo {author} {\bibfnamefont {R.}~\bibnamefont
  {Kubo}}\ and\ \bibinfo {author} {\bibfnamefont {K.}~\bibnamefont {Tomita}},\
  }\href {\doibase 10.1143/JPSJ.9.888} {\bibfield  {journal} {\bibinfo
  {journal} {J. Phys. Soc. Jpn.}\ }\textbf {\bibinfo {volume} {9}},\ \bibinfo
  {pages} {888} (\bibinfo {year} {1954})}\BibitemShut {NoStop}%
\bibitem [{\citenamefont {Mori}\ and\ \citenamefont
  {Kawasaki}(1962)}]{mori_esr}%
  \BibitemOpen
  \bibfield  {author} {\bibinfo {author} {\bibfnamefont {H.}~\bibnamefont
  {Mori}}\ and\ \bibinfo {author} {\bibfnamefont {K.}~\bibnamefont
  {Kawasaki}},\ }\href {\doibase 10.1143/PTP.28.971} {\bibfield  {journal}
  {\bibinfo  {journal} {Prog. Theor. Phys.}\ }\textbf {\bibinfo {volume}
  {28}},\ \bibinfo {pages} {971} (\bibinfo {year} {1962})}\BibitemShut
  {NoStop}%
\bibitem [{\citenamefont {Oshikawa}\ and\ \citenamefont
  {Affleck}(2002)}]{oshikawa_esr_spinchains}%
  \BibitemOpen
  \bibfield  {author} {\bibinfo {author} {\bibfnamefont {M.}~\bibnamefont
  {Oshikawa}}\ and\ \bibinfo {author} {\bibfnamefont {I.}~\bibnamefont
  {Affleck}},\ }\href@noop {} {\bibfield  {journal} {\bibinfo  {journal} {Phys.
  Rev. B}\ }\textbf {\bibinfo {volume} {65}},\ \bibinfo {pages} {134410}
  (\bibinfo {year} {2002})}\BibitemShut {NoStop}%
\bibitem [{\citenamefont {El~Shawish}\ \emph {et~al.}(2010)\citenamefont
  {El~Shawish}, \citenamefont {C\'epas},\ and\ \citenamefont
  {Miyashita}}]{ElShawish}%
  \BibitemOpen
  \bibfield  {author} {\bibinfo {author} {\bibfnamefont {S.}~\bibnamefont
  {El~Shawish}}, \bibinfo {author} {\bibfnamefont {O.}~\bibnamefont {C\'epas}},
  \ and\ \bibinfo {author} {\bibfnamefont {S.}~\bibnamefont {Miyashita}},\
  }\href {\doibase 10.1103/PhysRevB.81.224421} {\bibfield  {journal} {\bibinfo
  {journal} {Phys. Rev. B}\ }\textbf {\bibinfo {volume} {81}},\ \bibinfo
  {pages} {224421} (\bibinfo {year} {2010})}\BibitemShut {NoStop}%
\bibitem [{\citenamefont {Affleck}(1992)}]{affleck_esr_NENP}%
  \BibitemOpen
  \bibfield  {author} {\bibinfo {author} {\bibfnamefont {I.}~\bibnamefont
  {Affleck}},\ }\href {\doibase 10.1103/PhysRevB.46.9002} {\bibfield  {journal}
  {\bibinfo  {journal} {Phys. Rev. B}\ }\textbf {\bibinfo {volume} {46}},\
  \bibinfo {pages} {9002} (\bibinfo {year} {1992})}\BibitemShut {NoStop}%
\bibitem [{\citenamefont {Sakai}\ and\ \citenamefont
  {Shiba}(1994)}]{sakai_NENP}%
  \BibitemOpen
  \bibfield  {author} {\bibinfo {author} {\bibfnamefont {T.}~\bibnamefont
  {Sakai}}\ and\ \bibinfo {author} {\bibfnamefont {H.}~\bibnamefont {Shiba}},\
  }\href {\doibase 10.1143/JPSJ.63.867} {\bibfield  {journal} {\bibinfo
  {journal} {J. Phys. Soc. Jpn.}\ }\textbf {\bibinfo {volume} {63}},\ \bibinfo
  {pages} {867} (\bibinfo {year} {1994})}\BibitemShut {NoStop}%
\bibitem [{\citenamefont {Zvyagin}(2009)}]{zvyagin_esr}%
  \BibitemOpen
  \bibfield  {author} {\bibinfo {author} {\bibfnamefont {A.~A.}\ \bibnamefont
  {Zvyagin}},\ }\href {\doibase 10.1103/PhysRevB.79.064422} {\bibfield
  {journal} {\bibinfo  {journal} {Phys. Rev. B}\ }\textbf {\bibinfo {volume}
  {79}},\ \bibinfo {pages} {064422} (\bibinfo {year} {2009})}\BibitemShut
  {NoStop}%
\bibitem [{\citenamefont {\v{C}i\v{z}m\'ar}\ \emph {et~al.}(2010)\citenamefont
  {\v{C}i\v{z}m\'ar}, \citenamefont {Ozerov}, \citenamefont {Wosnitza},
  \citenamefont {Thielemann}, \citenamefont {Kr\"amer}, \citenamefont
  {R\"uegg}, \citenamefont {Piovesana}, \citenamefont {Klanj\v{s}ek},
  \citenamefont {Horvati\'{c}}, \citenamefont {Berthier},\ and\ \citenamefont
  {Zvyagin}}]{cizmar_esr_bpcb}%
  \BibitemOpen
  \bibfield  {author} {\bibinfo {author} {\bibfnamefont {E.}~\bibnamefont
  {\v{C}i\v{z}m\'ar}}, \bibinfo {author} {\bibfnamefont {M.}~\bibnamefont
  {Ozerov}}, \bibinfo {author} {\bibfnamefont {J.}~\bibnamefont {Wosnitza}},
  \bibinfo {author} {\bibfnamefont {B.}~\bibnamefont {Thielemann}}, \bibinfo
  {author} {\bibfnamefont {K.~W.}\ \bibnamefont {Kr\"amer}}, \bibinfo {author}
  {\bibfnamefont {C.}~\bibnamefont {R\"uegg}}, \bibinfo {author} {\bibfnamefont
  {O.}~\bibnamefont {Piovesana}}, \bibinfo {author} {\bibfnamefont
  {M.}~\bibnamefont {Klanj\v{s}ek}}, \bibinfo {author} {\bibfnamefont
  {M.}~\bibnamefont {Horvati\'{c}}}, \bibinfo {author} {\bibfnamefont
  {C.}~\bibnamefont {Berthier}}, \ and\ \bibinfo {author} {\bibfnamefont
  {S.~A.}\ \bibnamefont {Zvyagin}},\ }\href {\doibase
  10.1103/PhysRevB.82.054431} {\bibfield  {journal} {\bibinfo  {journal} {Phys.
  Rev. B}\ }\textbf {\bibinfo {volume} {82}},\ \bibinfo {pages} {054431}
  (\bibinfo {year} {2010})}\BibitemShut {NoStop}%
\bibitem [{\citenamefont {Feiguin}\ and\ \citenamefont
  {White}(2005)}]{White_finT}%
  \BibitemOpen
  \bibfield  {author} {\bibinfo {author} {\bibfnamefont {A.~E.}\ \bibnamefont
  {Feiguin}}\ and\ \bibinfo {author} {\bibfnamefont {S.~R.}\ \bibnamefont
  {White}},\ }\href@noop {} {\bibfield  {journal} {\bibinfo  {journal} {Phys.
  Rev. B}\ }\textbf {\bibinfo {volume} {72}},\ \bibinfo {pages} {220401}
  (\bibinfo {year} {2005})}\BibitemShut {NoStop}%
\bibitem [{\citenamefont {Bouillot}\ \emph {et~al.}(2011)\citenamefont
  {Bouillot}, \citenamefont {Kollath}, \citenamefont {L\"auchli}, \citenamefont
  {Zvonarev}, \citenamefont {Thielemann}, \citenamefont {R\"uegg},
  \citenamefont {Orignac}, \citenamefont {Citro}, \citenamefont
  {Klanj\ifmmode~\check{s}\else \v{s}\fi{}ek}, \citenamefont {Berthier},
  \citenamefont {Horvati\ifmmode~\acute{c}\else \'{c}\fi{}},\ and\
  \citenamefont {Giamarchi}}]{Bouillot_ladder_statics_dynamics}%
  \BibitemOpen
  \bibfield  {author} {\bibinfo {author} {\bibfnamefont {P.}~\bibnamefont
  {Bouillot}}, \bibinfo {author} {\bibfnamefont {C.}~\bibnamefont {Kollath}},
  \bibinfo {author} {\bibfnamefont {A.~M.}\ \bibnamefont {L\"auchli}}, \bibinfo
  {author} {\bibfnamefont {M.}~\bibnamefont {Zvonarev}}, \bibinfo {author}
  {\bibfnamefont {B.}~\bibnamefont {Thielemann}}, \bibinfo {author}
  {\bibfnamefont {C.}~\bibnamefont {R\"uegg}}, \bibinfo {author} {\bibfnamefont
  {E.}~\bibnamefont {Orignac}}, \bibinfo {author} {\bibfnamefont
  {R.}~\bibnamefont {Citro}}, \bibinfo {author} {\bibfnamefont
  {M.}~\bibnamefont {Klanj\ifmmode~\check{s}\else \v{s}\fi{}ek}}, \bibinfo
  {author} {\bibfnamefont {C.}~\bibnamefont {Berthier}}, \bibinfo {author}
  {\bibfnamefont {M.}~\bibnamefont {Horvati\ifmmode~\acute{c}\else
  \'{c}\fi{}}}, \ and\ \bibinfo {author} {\bibfnamefont {T.}~\bibnamefont
  {Giamarchi}},\ }\href {\doibase 10.1103/PhysRevB.83.054407} {\bibfield
  {journal} {\bibinfo  {journal} {Phys. Rev. B}\ }\textbf {\bibinfo {volume}
  {83}},\ \bibinfo {pages} {054407} (\bibinfo {year} {2011})}\BibitemShut
  {NoStop}%
\bibitem [{\citenamefont {Azzouz}(2006)}]{azzouz_short_correlation_ladder}%
  \BibitemOpen
  \bibfield  {author} {\bibinfo {author} {\bibfnamefont {M.}~\bibnamefont
  {Azzouz}},\ }\href {\doibase 10.1103/PhysRevB.74.174422} {\bibfield
  {journal} {\bibinfo  {journal} {Phys. Rev. B}\ }\textbf {\bibinfo {volume}
  {74}},\ \bibinfo {pages} {174422} (\bibinfo {year} {2006})}\BibitemShut
  {NoStop}%
\bibitem [{\citenamefont {Maeda}\ \emph {et~al.}(2005)\citenamefont {Maeda},
  \citenamefont {Sakai},\ and\ \citenamefont {Oshikawa}}]{Maeda_BetheAnsatz}%
  \BibitemOpen
  \bibfield  {author} {\bibinfo {author} {\bibfnamefont {Y.}~\bibnamefont
  {Maeda}}, \bibinfo {author} {\bibfnamefont {K.}~\bibnamefont {Sakai}}, \ and\
  \bibinfo {author} {\bibfnamefont {M.}~\bibnamefont {Oshikawa}},\ }\href
  {\doibase 10.1103/PhysRevLett.95.037602} {\bibfield  {journal} {\bibinfo
  {journal} {Phys. Rev. Lett.}\ }\textbf {\bibinfo {volume} {95}},\ \bibinfo
  {pages} {037602} (\bibinfo {year} {2005})}\BibitemShut {NoStop}%
\bibitem [{\citenamefont {Chitra}\ and\ \citenamefont
  {Giamarchi}(1997)}]{chitra_spinchains_field}%
  \BibitemOpen
  \bibfield  {author} {\bibinfo {author} {\bibfnamefont {R.}~\bibnamefont
  {Chitra}}\ and\ \bibinfo {author} {\bibfnamefont {T.}~\bibnamefont
  {Giamarchi}},\ }\href@noop {} {\bibfield  {journal} {\bibinfo  {journal}
  {Phys. Rev. B}\ }\textbf {\bibinfo {volume} {55}},\ \bibinfo {pages} {5816}
  (\bibinfo {year} {1997})}\BibitemShut {NoStop}%
\bibitem [{\citenamefont {Tachiki}\ \emph {et~al.}(1970)\citenamefont
  {Tachiki}, \citenamefont {Yamada},\ and\ \citenamefont
  {Maekawa}}]{Tachiki_spin_ladder}%
  \BibitemOpen
  \bibfield  {author} {\bibinfo {author} {\bibfnamefont {M.}~\bibnamefont
  {Tachiki}}, \bibinfo {author} {\bibfnamefont {T.}~\bibnamefont {Yamada}}, \
  and\ \bibinfo {author} {\bibfnamefont {S.}~\bibnamefont {Maekawa}},\
  }\href@noop {} {\bibfield  {journal} {\bibinfo  {journal} {J. Phys. Soc.
  Jpn.}\ }\textbf {\bibinfo {volume} {29}},\ \bibinfo {pages} {663} (\bibinfo
  {year} {1970})}\BibitemShut {NoStop}%
\bibitem [{\citenamefont {Chaboussant}\ \emph {et~al.}(1998)\citenamefont
  {Chaboussant}, \citenamefont {Julien}, \citenamefont {Fagot-Revurat},
  \citenamefont {Hanson}, \citenamefont {{L{\'e}vy}}, \citenamefont {Berthier},
  \citenamefont {{Horvati\'c}},\ and\ \citenamefont
  {Piovesana}}]{chaboussant_ladder_strongcoupling}%
  \BibitemOpen
  \bibfield  {author} {\bibinfo {author} {\bibfnamefont {G.}~\bibnamefont
  {Chaboussant}}, \bibinfo {author} {\bibfnamefont {M.-H.}\ \bibnamefont
  {Julien}}, \bibinfo {author} {\bibfnamefont {Y.}~\bibnamefont
  {Fagot-Revurat}}, \bibinfo {author} {\bibfnamefont {M.~E.}\ \bibnamefont
  {Hanson}}, \bibinfo {author} {\bibfnamefont {L.~P.}\ \bibnamefont
  {{L{\'e}vy}}}, \bibinfo {author} {\bibfnamefont {C.}~\bibnamefont
  {Berthier}}, \bibinfo {author} {\bibfnamefont {M.}~\bibnamefont
  {{Horvati\'c}}}, \ and\ \bibinfo {author} {\bibfnamefont {O.}~\bibnamefont
  {Piovesana}},\ }\href@noop {} {\bibfield  {journal} {\bibinfo  {journal}
  {Eur. Phys. J. B}\ }\textbf {\bibinfo {volume} {6}},\ \bibinfo {pages} {167}
  (\bibinfo {year} {1998})}\BibitemShut {NoStop}%
\bibitem [{\citenamefont {Mila}(1998)}]{mila_ladder_strongcoupling}%
  \BibitemOpen
  \bibfield  {author} {\bibinfo {author} {\bibfnamefont {F.}~\bibnamefont
  {Mila}},\ }\href@noop {} {\bibfield  {journal} {\bibinfo  {journal} {Eur.
  Phys. J. B}\ }\textbf {\bibinfo {volume} {6}},\ \bibinfo {pages} {201}
  (\bibinfo {year} {1998})}\BibitemShut {NoStop}%
\bibitem [{\citenamefont {Giamarchi}\ and\ \citenamefont
  {Tsvelik}(1999)}]{giamarchi_ladder_coupled}%
  \BibitemOpen
  \bibfield  {author} {\bibinfo {author} {\bibfnamefont {T.}~\bibnamefont
  {Giamarchi}}\ and\ \bibinfo {author} {\bibfnamefont {A.~M.}\ \bibnamefont
  {Tsvelik}},\ }\href@noop {} {\bibfield  {journal} {\bibinfo  {journal} {Phys.
  Rev. B}\ }\textbf {\bibinfo {volume} {59}},\ \bibinfo {pages} {11398}
  (\bibinfo {year} {1999})}\BibitemShut {NoStop}%
\bibitem [{\citenamefont {Patyal}\ \emph {et~al.}(1990)\citenamefont {Patyal},
  \citenamefont {Scott},\ and\ \citenamefont {Willett}}]{Patyal_BPCB}%
  \BibitemOpen
  \bibfield  {author} {\bibinfo {author} {\bibfnamefont {B.~R.}\ \bibnamefont
  {Patyal}}, \bibinfo {author} {\bibfnamefont {B.~L.}\ \bibnamefont {Scott}}, \
  and\ \bibinfo {author} {\bibfnamefont {R.~D.}\ \bibnamefont {Willett}},\
  }\href@noop {} {\bibfield  {journal} {\bibinfo  {journal} {Phys. Rev. B}\
  }\textbf {\bibinfo {volume} {41}},\ \bibinfo {pages} {1657} (\bibinfo {year}
  {1990})}\BibitemShut {NoStop}%
\bibitem [{\citenamefont {Watson}\ \emph {et~al.}(2001)\citenamefont {Watson},
  \citenamefont {Kotov}, \citenamefont {Meisel}, \citenamefont {Hall},
  \citenamefont {Granroth}, \citenamefont {Montfrooij}, \citenamefont {Nagler},
  \citenamefont {Jensen}, \citenamefont {Backov}, \citenamefont {Petruska},
  \citenamefont {Fanucci},\ and\ \citenamefont {Talham}}]{watson_bpcb}%
  \BibitemOpen
  \bibfield  {author} {\bibinfo {author} {\bibfnamefont {B.~C.}\ \bibnamefont
  {Watson}}, \bibinfo {author} {\bibfnamefont {V.~N.}\ \bibnamefont {Kotov}},
  \bibinfo {author} {\bibfnamefont {M.~W.}\ \bibnamefont {Meisel}}, \bibinfo
  {author} {\bibfnamefont {D.~W.}\ \bibnamefont {Hall}}, \bibinfo {author}
  {\bibfnamefont {G.~E.}\ \bibnamefont {Granroth}}, \bibinfo {author}
  {\bibfnamefont {W.~T.}\ \bibnamefont {Montfrooij}}, \bibinfo {author}
  {\bibfnamefont {S.~E.}\ \bibnamefont {Nagler}}, \bibinfo {author}
  {\bibfnamefont {D.~A.}\ \bibnamefont {Jensen}}, \bibinfo {author}
  {\bibfnamefont {R.}~\bibnamefont {Backov}}, \bibinfo {author} {\bibfnamefont
  {M.~A.}\ \bibnamefont {Petruska}}, \bibinfo {author} {\bibfnamefont {G.~E.}\
  \bibnamefont {Fanucci}}, \ and\ \bibinfo {author} {\bibfnamefont {D.~R.}\
  \bibnamefont {Talham}},\ }\href@noop {} {\bibfield  {journal} {\bibinfo
  {journal} {Phys. Rev. Lett.}\ }\textbf {\bibinfo {volume} {86}},\ \bibinfo
  {pages} {5168} (\bibinfo {year} {2001})}\BibitemShut {NoStop}%
\bibitem [{\citenamefont {Thielemann}\ \emph
  {et~al.}(2009{\natexlab{a}})\citenamefont {Thielemann}, \citenamefont
  {R\"uegg}, \citenamefont {Kiefer}, \citenamefont {R\o{}nnow}, \citenamefont
  {Normand}, \citenamefont {Bouillot}, \citenamefont {Kollath}, \citenamefont
  {Orignac}, \citenamefont {Citro}, \citenamefont {Giamarchi}, \citenamefont
  {L\"auchli}, \citenamefont {Biner}, \citenamefont {Kr\"amer}, \citenamefont
  {Wolff-Fabris}, \citenamefont {Zapf}, \citenamefont {Jaime}, \citenamefont
  {Stahn}, \citenamefont {Christensen}, \citenamefont {Grenier}, \citenamefont
  {McMorrow},\ and\ \citenamefont {Mesot}}]{Thielemann_ND_3Dladder}%
  \BibitemOpen
  \bibfield  {author} {\bibinfo {author} {\bibfnamefont {B.}~\bibnamefont
  {Thielemann}}, \bibinfo {author} {\bibfnamefont {C.}~\bibnamefont {R\"uegg}},
  \bibinfo {author} {\bibfnamefont {K.}~\bibnamefont {Kiefer}}, \bibinfo
  {author} {\bibfnamefont {H.~M.}\ \bibnamefont {R\o{}nnow}}, \bibinfo {author}
  {\bibfnamefont {B.}~\bibnamefont {Normand}}, \bibinfo {author} {\bibfnamefont
  {P.}~\bibnamefont {Bouillot}}, \bibinfo {author} {\bibfnamefont
  {C.}~\bibnamefont {Kollath}}, \bibinfo {author} {\bibfnamefont
  {E.}~\bibnamefont {Orignac}}, \bibinfo {author} {\bibfnamefont
  {R.}~\bibnamefont {Citro}}, \bibinfo {author} {\bibfnamefont
  {T.}~\bibnamefont {Giamarchi}}, \bibinfo {author} {\bibfnamefont {A.~M.}\
  \bibnamefont {L\"auchli}}, \bibinfo {author} {\bibfnamefont {D.}~\bibnamefont
  {Biner}}, \bibinfo {author} {\bibfnamefont {K.~W.}\ \bibnamefont {Kr\"amer}},
  \bibinfo {author} {\bibfnamefont {F.}~\bibnamefont {Wolff-Fabris}}, \bibinfo
  {author} {\bibfnamefont {V.~S.}\ \bibnamefont {Zapf}}, \bibinfo {author}
  {\bibfnamefont {M.}~\bibnamefont {Jaime}}, \bibinfo {author} {\bibfnamefont
  {J.}~\bibnamefont {Stahn}}, \bibinfo {author} {\bibfnamefont {N.~B.}\
  \bibnamefont {Christensen}}, \bibinfo {author} {\bibfnamefont
  {B.}~\bibnamefont {Grenier}}, \bibinfo {author} {\bibfnamefont {D.~F.}\
  \bibnamefont {McMorrow}}, \ and\ \bibinfo {author} {\bibfnamefont
  {J.}~\bibnamefont {Mesot}},\ }\href@noop {} {\bibfield  {journal} {\bibinfo
  {journal} {Phys. Rev. B}\ }\textbf {\bibinfo {volume} {79}},\ \bibinfo
  {pages} {020408(R)} (\bibinfo {year} {2009}{\natexlab{a}})}\BibitemShut
  {NoStop}%
\bibitem [{\citenamefont {Thielemann}\ \emph
  {et~al.}(2009{\natexlab{b}})\citenamefont {Thielemann}, \citenamefont
  {R\"uegg}, \citenamefont {R\o{}nnow}, \citenamefont {L\"auchli},
  \citenamefont {Caux}, \citenamefont {Normand}, \citenamefont {Biner},
  \citenamefont {Kr\"amer}, \citenamefont {G\"udel}, \citenamefont {Stahn},
  \citenamefont {Habicht}, \citenamefont {Kiefer}, \citenamefont {Boehm},
  \citenamefont {McMorrow},\ and\ \citenamefont
  {Mesot}}]{Thielemann_INS_ladder}%
  \BibitemOpen
  \bibfield  {author} {\bibinfo {author} {\bibfnamefont {B.}~\bibnamefont
  {Thielemann}}, \bibinfo {author} {\bibfnamefont {C.}~\bibnamefont {R\"uegg}},
  \bibinfo {author} {\bibfnamefont {H.~M.}\ \bibnamefont {R\o{}nnow}}, \bibinfo
  {author} {\bibfnamefont {A.~M.}\ \bibnamefont {L\"auchli}}, \bibinfo {author}
  {\bibfnamefont {J.-S.}\ \bibnamefont {Caux}}, \bibinfo {author}
  {\bibfnamefont {B.}~\bibnamefont {Normand}}, \bibinfo {author} {\bibfnamefont
  {D.}~\bibnamefont {Biner}}, \bibinfo {author} {\bibfnamefont {K.~W.}\
  \bibnamefont {Kr\"amer}}, \bibinfo {author} {\bibfnamefont {H.-U.}\
  \bibnamefont {G\"udel}}, \bibinfo {author} {\bibfnamefont {J.}~\bibnamefont
  {Stahn}}, \bibinfo {author} {\bibfnamefont {K.}~\bibnamefont {Habicht}},
  \bibinfo {author} {\bibfnamefont {K.}~\bibnamefont {Kiefer}}, \bibinfo
  {author} {\bibfnamefont {M.}~\bibnamefont {Boehm}}, \bibinfo {author}
  {\bibfnamefont {D.~F.}\ \bibnamefont {McMorrow}}, \ and\ \bibinfo {author}
  {\bibfnamefont {J.}~\bibnamefont {Mesot}},\ }\href@noop {} {\bibfield
  {journal} {\bibinfo  {journal} {Phys. Rev. Lett.}\ }\textbf {\bibinfo
  {volume} {102}},\ \bibinfo {pages} {107204} (\bibinfo {year}
  {2009}{\natexlab{b}})}\BibitemShut {NoStop}%
\bibitem [{\citenamefont {Savici}\ \emph {et~al.}(2009)\citenamefont {Savici},
  \citenamefont {Granroth}, \citenamefont {Broholm}, \citenamefont
  {Pajerowski}, \citenamefont {Brown}, \citenamefont {Talham}, \citenamefont
  {Meisel}, \citenamefont {Schmidt}, \citenamefont {Uhrig},\ and\ \citenamefont
  {Nagler}}]{Savici_BPCB_INS}%
  \BibitemOpen
  \bibfield  {author} {\bibinfo {author} {\bibfnamefont {A.~T.}\ \bibnamefont
  {Savici}}, \bibinfo {author} {\bibfnamefont {G.~E.}\ \bibnamefont
  {Granroth}}, \bibinfo {author} {\bibfnamefont {C.~L.}\ \bibnamefont
  {Broholm}}, \bibinfo {author} {\bibfnamefont {D.~M.}\ \bibnamefont
  {Pajerowski}}, \bibinfo {author} {\bibfnamefont {C.~M.}\ \bibnamefont
  {Brown}}, \bibinfo {author} {\bibfnamefont {D.~R.}\ \bibnamefont {Talham}},
  \bibinfo {author} {\bibfnamefont {M.~W.}\ \bibnamefont {Meisel}}, \bibinfo
  {author} {\bibfnamefont {K.~P.}\ \bibnamefont {Schmidt}}, \bibinfo {author}
  {\bibfnamefont {G.~S.}\ \bibnamefont {Uhrig}}, \ and\ \bibinfo {author}
  {\bibfnamefont {S.~E.}\ \bibnamefont {Nagler}},\ }\href@noop {} {\bibfield
  {journal} {\bibinfo  {journal} {Phys. Rev. B}\ }\textbf {\bibinfo {volume}
  {80}},\ \bibinfo {pages} {094411} (\bibinfo {year} {2009})}\BibitemShut
  {NoStop}%
\bibitem [{\citenamefont {R\"uegg}\ \emph {et~al.}(2008)\citenamefont
  {R\"uegg}, \citenamefont {Kiefer}, \citenamefont {Thielemann}, \citenamefont
  {McMorrow}, \citenamefont {Zapf}, \citenamefont {Normand}, \citenamefont
  {Zvonarev}, \citenamefont {Bouillot}, \citenamefont {Kollath}, \citenamefont
  {Giamarchi}, \citenamefont {Capponi}, \citenamefont {Poilblanc},
  \citenamefont {Biner},\ and\ \citenamefont {Kr\"amer}}]{Ruegg_thermo_ladder}%
  \BibitemOpen
  \bibfield  {author} {\bibinfo {author} {\bibfnamefont {C.}~\bibnamefont
  {R\"uegg}}, \bibinfo {author} {\bibfnamefont {K.}~\bibnamefont {Kiefer}},
  \bibinfo {author} {\bibfnamefont {B.}~\bibnamefont {Thielemann}}, \bibinfo
  {author} {\bibfnamefont {D.~F.}\ \bibnamefont {McMorrow}}, \bibinfo {author}
  {\bibfnamefont {V.}~\bibnamefont {Zapf}}, \bibinfo {author} {\bibfnamefont
  {B.}~\bibnamefont {Normand}}, \bibinfo {author} {\bibfnamefont {M.~B.}\
  \bibnamefont {Zvonarev}}, \bibinfo {author} {\bibfnamefont {P.}~\bibnamefont
  {Bouillot}}, \bibinfo {author} {\bibfnamefont {C.}~\bibnamefont {Kollath}},
  \bibinfo {author} {\bibfnamefont {T.}~\bibnamefont {Giamarchi}}, \bibinfo
  {author} {\bibfnamefont {S.}~\bibnamefont {Capponi}}, \bibinfo {author}
  {\bibfnamefont {D.}~\bibnamefont {Poilblanc}}, \bibinfo {author}
  {\bibfnamefont {D.}~\bibnamefont {Biner}}, \ and\ \bibinfo {author}
  {\bibfnamefont {K.~W.}\ \bibnamefont {Kr\"amer}},\ }\href@noop {} {\bibfield
  {journal} {\bibinfo  {journal} {Phys. Rev. Lett.}\ }\textbf {\bibinfo
  {volume} {101}},\ \bibinfo {pages} {247202} (\bibinfo {year}
  {2008})}\BibitemShut {NoStop}%
\bibitem [{\citenamefont {Anfuso}\ \emph {et~al.}(2008)\citenamefont {Anfuso},
  \citenamefont {Garst}, \citenamefont {Rosch}, \citenamefont {Heyer},
  \citenamefont {Lorenz}, \citenamefont {R\"uegg},\ and\ \citenamefont
  {Kr\"amer}}]{Anfuso_BPCB_magnetostriction}%
  \BibitemOpen
  \bibfield  {author} {\bibinfo {author} {\bibfnamefont {F.}~\bibnamefont
  {Anfuso}}, \bibinfo {author} {\bibfnamefont {M.}~\bibnamefont {Garst}},
  \bibinfo {author} {\bibfnamefont {A.}~\bibnamefont {Rosch}}, \bibinfo
  {author} {\bibfnamefont {O.}~\bibnamefont {Heyer}}, \bibinfo {author}
  {\bibfnamefont {T.}~\bibnamefont {Lorenz}}, \bibinfo {author} {\bibfnamefont
  {C.}~\bibnamefont {R\"uegg}}, \ and\ \bibinfo {author} {\bibfnamefont
  {K.}~\bibnamefont {Kr\"amer}},\ }\href@noop {} {\bibfield  {journal}
  {\bibinfo  {journal} {Phys. Rev. B}\ }\textbf {\bibinfo {volume} {77}},\
  \bibinfo {pages} {235113} (\bibinfo {year} {2008})}\BibitemShut {NoStop}%
\bibitem [{\citenamefont {Lorenz}\ \emph {et~al.}(2008)\citenamefont {Lorenz},
  \citenamefont {Heyer}, \citenamefont {Garst}, \citenamefont {Anfuso},
  \citenamefont {Rosch}, \citenamefont {R\"uegg},\ and\ \citenamefont
  {Kr\"amer}}]{lorenz_thermalexp_magnetostriction}%
  \BibitemOpen
  \bibfield  {author} {\bibinfo {author} {\bibfnamefont {T.}~\bibnamefont
  {Lorenz}}, \bibinfo {author} {\bibfnamefont {O.}~\bibnamefont {Heyer}},
  \bibinfo {author} {\bibfnamefont {M.}~\bibnamefont {Garst}}, \bibinfo
  {author} {\bibfnamefont {F.}~\bibnamefont {Anfuso}}, \bibinfo {author}
  {\bibfnamefont {A.}~\bibnamefont {Rosch}}, \bibinfo {author} {\bibfnamefont
  {C.}~\bibnamefont {R\"uegg}}, \ and\ \bibinfo {author} {\bibfnamefont
  {K.}~\bibnamefont {Kr\"amer}},\ }\href@noop {} {\bibfield  {journal}
  {\bibinfo  {journal} {Phys. Rev. Lett.}\ }\textbf {\bibinfo {volume} {100}},\
  \bibinfo {pages} {067208} (\bibinfo {year} {2008})}\BibitemShut {NoStop}%
\end{thebibliography}
\end{document}